\documentclass{aa}
\usepackage{natbib}
\usepackage{xspace}
\usepackage{amssymb}
\usepackage{amsmath}
\usepackage{graphicx}
\usepackage{comment}

\def\aap{A\& A}

\def\mnras{MNRAS}
\def\nat{Nature}
\def\apj{ApJ}

\def\apjl{ApJL}
\def\apjs{ApJS}

\def\um{\ensuremath{\mu\mathrm{m}}\xspace}

\def\rcentr{\ensuremath{r_\mathrm{centr}}\xspace}
\def\ftr{\ensuremath{f_\mathrm{trickle}}\xspace}
\begin{document}
\title{Coagulation of small grains in disks: the influence of residual 
 infall and initial small-grain content}
\titlerunning{Coagulation of small grains in disks}
\authorrunning{Dominik, Dullemond}
\author{C.~Dominik$^{1,2}$ \& C.P.~Dullemond$^{3}$}
\institute{$^{1}$Sterrenkundig Instituut `Anton Pannekoek', Kruislaan 403,
NL-1098 SJ Amsterdam, The Netherlands; E--mail:
dominik@science.uva.nl\\
$^{2}$Afdeling Sterrenkunde, Radboud Universiteit Nijmegen,
Postbus 9010, 6500 GL Nijmegen\\
$^{3}$Max Planck Institut f\"ur Astronomie, K\"onigstuhl 17,
D-69117 Heidelberg, Germany; e--mail: dullemon@mpia.de}
\date{DRAFT, \today} 

\abstract{Turbulent coagulation in protoplanetary disks is known to
  operate on timescale far shorter than the lifetime of the
  disk. In the absence of mechanisms that replenish the small
    dust grain population, protoplanetary disks would rapidly lose
    their continuum opacity-bearing dust.  This is inconsistent with
  infrared observations of disks around T Tauri stars and Herbig Ae/Be
  stars, which are usually optically thick at visual wavelengths and
  show signatures of small ($a\lesssim 3\mu$m) grains. A plausible
  replenishing mechanism of small grains is collisional fragmentation
  or erosion of large dust aggregates, which model calculations
  predict to play an important role in protoplanetary disks.  If
  optically thick disks are to be seen as proof for ongoing
  fragmentation or erosion, then alternative explanations for the
  existence of optically thick disks must be studied carefully.  In
  tis study we explore two scenarios.  First, we study the effect of
  residual, low-level infall of matter onto the disk surface. We find
  that infall rates as low as 10$^{-11}$M$_{\odot}$/yr can, in
  principle, replenish the small grain population to a level that
  keeps the disk marginally optically thick.  However, it remains to
  be seen if the assumption of such inflow is realistic for star+disk
  systems at the age of several Myrs, at which winds and jets are
  expected to have removed any residual envelope.  The effectiveness
  of even a low level infall can be understood by the strongly
  non-linear behavior of the coagulation equation: a high, fine-grain,
  dust density at any given time leads to very, effective removal of
  these small grains, while a low fine-grain density lasts for a far
  longer time.  We then consider a second scenario in which, during
  the buildup phase of the disk, an intermediate fine-grain dust
  abundance is generated that is sufficiently low to ensure longevity
  of the state yet sufficiently high for the disk to remain optically
  thick.  While our models confirm that such an ``initial condition'' can be
  constructed, we argue that these special initial conditions
  cannot be achieved during the disk build-up phase.  In summary,
  fragmentation or erosion still appear to be the most promising
processes to explain the abundant presence of small grains in old
disks.}

\maketitle


\section{Introduction}
The first stages of planet formation are now directly observable with
infrared observations of protoplanetary disks. Most studies focus on
the presence of slightly grown dust grains as evidence of the onset of
this growth process \citep[e.g.][]{2003A&A...400L..21V,
  2003A&A...412L..43P, 2004Natur.432..479V,2006ApJ...639..275K}.
However, a much undervalued fact is that even old ($\sim $3 Myr or
even older) disks still contain significant populations of small
($\lesssim 3\mu$m) grains as indicated by the presence of strong
infrared solid state emission features \citep{1998A&A...332L..25M} and
large optical depth \citep{2003A&A...401..577B}.  By optically thick,
we refer to a significant optical depth as measured in the V band,
which causes the disk to absorb or scatter all incident stellar flux
Models of dust coagulation without fragmentation predict consistently
severe depletion in these small grains on timescale far smaller than
the disk lifetime, as seen in calculations by
\citet{1984Icar...60..553W}, \citet[hereafter
DD05]{2005A&A...434..971D}, and \citet[hereafter
TIH05]{2005ApJ...625..414T}.  While the assumption of a perfectly
laminar disk may allow a small population of micron-size grains to
survive (TIH05), even very weak turbulence (at a level of
$\alpha=10^{-6}$) would enhance the coagulation process sufficiently
for the disk to become optically thin.  In the above models,
differential radial drift was not included: at low levels of
turbulence, it is the dominant mechanism controlling the relative
velocities of grains of various sizes
\citep{1972fpp..conf..211W,1977MNRAS.180...57W,1997Icar..127..290W,2008A&A...480..859B},
which implies that, without fragmentation, growth is enhanced by this
mechanism. 

  Models of dust coagulation without a treatment of
  collisional fragmentation are of course
  unrealistic. \citet{1984Icar...60..553W,1997Icar..127..290W}
  included fragmentation in his model.  It turns out  that
  fragmentation is a universal outcome.  Models of particle
  motion in laminar and turbulent accretion disks predict collision
  velocities in excess of the critical value for break-up
  (Weidenschilling 1977; V\"olk et al.~1984; Blum \& Wurm 2000; for a
  review, see Dominik et al.~(2007))\nocite{2000Icar..143..138B,
    1977MNRAS.180...57W, 2007prpl.conf..783D}.
  \citet{2008A&A...480..859B} demonstrated that collisional
  fragmentation is a major barrier to the growth of dust aggregates
  beyond centimeter or decimeter sizes.  
However, it is still unknown if fragmentation can produce a sufficient
amount of \emph{small} grain dust to produce the disk opacity inferred from
observations. 
  This
  question is difficult to answer, because the amount of small grains
  produced in collisions is strongly dependent on collision parameters
  (Paszun and Dominik, submitted to A\&A).  It is therefore important
  to know if the opaqueness of protoplanetary disks is an indication
  of small grain production by fragmentation and erosion (as we
  argued in DD05), or if other mechanisms could retain the small grain
  population in disks, such as the thermal breakup of icy aggregates
  (Tanaka et al.~2005, poster at PPV), which can operate in a
  limited radial range.
  
  In the present paper, we investigate if a continuous
  replenishment of small grains by residual infall of matter from
  the circumstellar envelope onto the disk's surface could
  sustain the small dust grain population.  This infall might
  be residual infall from the parental cloud, or 
  Bondi-Hoyle like accretion \citep{1944MNRAS.104..273B} by a
  protoplanetary disk moving through a molecular cloud.  We assume
  that the infall originates in the parental
  cloud, with zero velocity at infinity.  As long as the motion of the
  star relative to the cloud is small compared with Keplerian velocities
  in the inner regions of the disk, these cases are
  similar.  The idea of external replenishment was first studied by
\citet{1988A&A...195..183M}, who were interested primarily in
convection-driven turbulence (which requires optically-thick disks)
during the early stages of the disk's lifetime.  Using a simple
one-zone model of a disk, they indeed showed that a mass infall rate
of $10^{-6}$ and $10^{-7}M_{\odot}/$yr could help to keep the optical
depth of the disk high.

One can easily show that an envelope with an infall rate of
$10^{-6}M_{\odot}/$yr would be optically thick and completely obscure
the star+disk system, while 10$^{-7}$M$_{\odot}$/yr would still be
marginally optically thick.  The typical star+disk systems studied in
the context of grain growth and planet formation are `revealed
sources' (class II in the Lada classification), meaning that they have
envelopes that are optically thin.  If this process is to work at all
in these systems, it must work at relatively low infall rates, which
we consider here.

In Sect.~\ref{sec-results-1} we first study if the replenishment of
the disk with small dust grains by an infalling tenuous envelope with
low infall rate would be sufficient to ensure that the disk remained
sufficiently optically
thick to be consistent with observations.  We
employ the 1-D vertical slice models for coagulation described by
DD05, but with additional continuous infall of small grained dust at
the top.  We study the total vertical optical depth, the optical
depth in grains smaller than 3 $\mu$m (which are the
silicate-feature-producing grains), and the optical depth in grains
larger than 3 \um.

In Sect.~\ref{sec-results-2} we consider the effects
of a low initial abundance of small grains on the removal timescale of
small grains.  This involves models without external inflow, but with
different dust-to-gas ratios in the initial setup.  

We discuss our results and present our conclusions in
Sect.~\ref{sec-discuss}.

\begin{table*}[htbp]
  \caption[]{\label{tab:example_values} Values for various quantities
    at different distances $r$ from the star, for the model with
    trickling rate 10$^{-9}$M$_{\odot}$/yr.  The quantities are: 
    surface density $\Sigma$, trickling flux $\ftr$, impact velocity
    of the infall $v_{\theta}$, midplane temperature $T_{\rm mid}$,
    pressure scale height $H_p$, accretion shock height $H_{\rm sh}$,
    stopping time for 0.1\um grains $t_{\rm stop}$, deceleration length
    for 0.1\um grains in units of the shock height.}
\vspace{4mm}
\begin{center}
\begin{tabular}[tb]{rrrrrrrrr}
\multicolumn{1}{c}{r}    &
\multicolumn{1}{c}{$\Sigma$}    & 
\multicolumn{1}{c}{\ftr}  & 
\multicolumn{1}{c}{$v_{\theta}$} & 
\multicolumn{1}{c}{T$_{\rm mid}$} 
& \multicolumn{1}{c}{$H_p/r$} 
& \multicolumn{1}{c}{z$_{\rm sh}/r$}
& \multicolumn{1}{c}{t$_{\rm stop}$}
& \multicolumn{1}{c}{$\Delta z/H_{\rm sh}$} \\
\multicolumn{1}{c}{AU}   & 
\multicolumn{1}{c}{g cm$^{-2}$} 
& \multicolumn{1}{c}{g cm$^{-2}$ s$^{-2}$} 
& \multicolumn{1}{c}{km/s}
& \multicolumn{1}{c}{K}             &         &                & 
\multicolumn{1}{c}{days}           &                       \\
\hline
\hline
1 & 354 & 5.6\,(-14) & 21 & 305 & 0.05 & 5.83 & 3.7 & 0.12 \\
10 & 35 & 5.8\,(-15) & 6.5 & 96 & 0.09 & 5.3 & 204 & 0.12 \\
100 & 3.5 & 8.0\,(-16) & 1.5 & 30 & 0.16 & 4.7 & 14600 & 0.12 \\
\end{tabular}
\end{center}
\end{table*}

\section{The model}

\subsection{Disk and coagulation model}
We model the dust coagulation in a disk around a pre-main-sequence
star.  We begin the simulation at $t=0$, which represents the time
immediately after the main build-up phase of the star+disk
system. This is the moment when the main infall onto the system, which
is of the order of $10^{-5}M_{\odot}/$yr during the build-up phase,
either ceases entirely or reduces to far lower values. It is this
much lower `residual infall' that we focus on in
Sect.~\ref{sec-results-1}.

As our base star+disk model, we consider a 0.05M$_{\odot}$ disk around
a 0.5M$_{\odot}$ star.  The star is assumed to have an effective
temperature of 4000\,K and a radius of 2.5R$_{\odot}$.  For the surface
density of the disk, we assume a power law
$\Sigma=\Sigma_0(r/\mathrm{AU})^{-1}$ (see e.g. Dullemond et al 2007
for a review\nocite{2007prpl.conf..555D}) where $r$ is the distance
from the star. The disk radius is assumed to be 200 AU, and
$M_{\mathrm{disk}}=0.1M_{*}=0.05M_{\odot}$, such that the surface
density at 1 AU becomes $\Sigma_0=354$g/cm$^2$ (in gas). The vertical
gas distribution is assumed to be in hydrostatic equilibrium, with a
gas temperature given by a simple irradiation recipe
$T_{\mathrm{gas}}(r) = (1/\sqrt{2}) \phi^{1/4}T_{*}
\sqrt{R_{\star}/r}$, where $\phi$ is the irradiation angle, which we
assume to be $0.05$. The dust and gas are initially well-mixed, with a
dust-to-gas ratio by mass of 0.01.  The dust grain size distribution
in the initial disk model and in the infalling flow of matter
is assumed to be an MRN distribution \citep{1977ApJ...217..425M} with
particles distributed between 0.1 and 0.5\um, following a power-law size
distribution $f(a)\propto a^{-7/2}$.  The dust grains are assumed to
be astronomical silicate. The opacities were computed using Mie
theory, with optical properties from \citet{1993ApJ...402..441L}.

The coagulation model is that of DD05, which we used unmodified
for these simulations. The disk's gas density and temperature were
assumed to remain constant in spite of the vastly changing dust
properties with time. We included coagulation by Brownian motion,
vertical settling, and turbulent motions.  The calculations do
not include coagulation driven by radial drift in the disk.  Radial
drift is known to be important for the growth of large particles
\citep{1980Icar...44..172W,1984Icar...60..553W}.  In the current paper,
we are interested in the small grain component, for which
the differences are expected to be minor.  Furthermore, it allows us to
compare directly with the results by DD05, who emphasized a
need for some replenishment mechanism of small grains.  
In a model with strong turbulent mixing, the coagulation occurs
particularly rapidly as material from the disk upper layers is mixed
more deeply inside the disk, where the densities are higher and
coagulation timescales are far shorter. 

\subsection{Infalling envelope}
For the computations in this paper, we assume that the infalling
matter is the residual influx from the collapse of a rotating cloud.
The density and velocity distribution for this case were described
with analytical models by \citet{1976ApJ...210..377U}, representing a
simplified analysis of a the more general study by
\citet{1984ApJ...286..529T}.  The main parameters of this model are
the infall rate $\dot M$ and the centrifugal radius $\rcentr$, which
we fixed to be 200\,AU.  The gas density at a distance $r$ and polar
angle $\mu=\cos\theta$ is given by:
\begin{equation}
\rho(r,\theta) = \frac{\dot M}{4\pi \sqrt{GMr^3}}
\left(1+\frac{\mu}{\mu_0}\right)^{-1/2}\left(\frac{\mu}{\mu_0}
+\frac{2\mu_0^2 \rcentr}{r}\right)^{-1}
\end{equation}
where $\mu_0$ is a solution of the equation:
\begin{equation}
\frac{r}{\rcentr}\left(1-\frac{\mu}{\mu_0}\right)=1-\mu_0^2 \quad.
\end{equation}
The $r$- and $\theta$-velocity components are:
\begin{eqnarray}
v_r &=& -\left(\frac{GM}{r}\right)^{1/2}\left(1+\frac{\mu}{\mu_0}\right)^{1/2} \\
v_\theta &=& \left(\frac{GM}{r}\right)^{1/2} 
\left(\frac{\mu_0-\mu}{\sqrt{1-\mu^2}}\right)
\left(1+\frac{\mu}{\mu_0}\right)^{1/2} \quad.
\end{eqnarray}
Multiplying $v_\theta(\theta=\pi/2)$ by the density $\rho(\theta=\theta/2)$
provides the mass flux onto the disk surface:
\begin{equation}\label{eq-local-trickling-rate}
\ftr = \frac{\dot M}{8\pi r \mu_0 \rcentr}
\end{equation}
where $\mu=\sqrt{1-r/\rcentr}$.

Some example values for the trickling inflow onto the disk surface are
given in Table~\ref{tab:example_values}

\subsection{Location of the accretion shock and dust injection}

As matter falls onto the disk, an accretion shock must form to
decelerate the gas.  The physics of the accretion shock is complex
\citep[e.g.][]{1994Icar..112..430R}, because the temperature behind
the accretion shock has to be computed.  In the current study we
assume that the cooling behind the shock
is fast, i.e. that the post-shock gas quickly cools to the local
temperature of the disk.  We may then compute the location of the
shock by equating the ram pressure in the infalling material to the
local hydrostatic pressure in the disk and solve for the height above
the disk.  The ram pressure from the infalling material is given by
\begin{equation}
\label{eq:1}
p_{\rm ram}=\rho(r,0)\,v^2_\theta(r) \quad.
\end{equation}
The hydrostatic pressure in the disk $p_{\rm disk}$ is given as a function
of height $z$ in the disk by
\begin{equation}
\label{eq:2}
p_{\rm disk}=\gamma\rho_{\rm disk}c_s^2 = \frac{\Sigma\gamma
  c_s^2}{H_p \sqrt{2\pi}} \exp \left\{ - \frac{z^2}{2H_p^2} \right\}\quad,
\end{equation}
where $H_p=c_s/\Omega_k$ is the pressure scale height in the disk,
$c_s=\sqrt{kT/\mu m_p}$ is the sound velocity, $\mu$ is the
mean molecular weight, $m_p$ the proton mass,
$\Omega_k=\sqrt{GM_\star/r^3}$ is the local Kepler frequency, and
$\gamma$ is the adiabatic index.
\begin{equation}
\label{eq:3}
z_{\rm sh}=\sqrt{-2H_p\ln \left[ \frac{\sqrt{2\pi}H_pp_{\rm
        ram}}{\gamma c_s^{2}\Sigma} \right]} \quad.
\end{equation}

\noindent
We find that this is typically 5 to 6 pressure scales above the disk
(for detailed values, see Table~\ref{tab:example_values}).

When the infalling gas motion has stopped, the dust particles are
injected into the disk with the free-fall speed at the considered
location.  Dust grains are decelerated over a certain distance, which
we compute numerically.  The results are reported in Table
\ref{tab:example_values} and indicate that the stopping time is
approximately a few hours, and the stopping length is far smaller than
the disk scale height.  Another issue is the temperature of the grains
during deceleration.  The grains heat up due to frictional coupling
with the gas.  We can estimate the heating with simple argument.  The
kinetic energy of the incoming dust grain is given by $0.5m_{\rm
  gr}v_\theta^2$.  Following the deceleration curve of the grain, we
make the extreme assumption that the entire kinetic energy is
transformed into heating grains (in reality, a large fraction
will heat the surrounding gas).  Equating this heating rate to a
radiative cooling rate, we found that the grains were heated up by
less than a 
few hundred degrees at 1AU, and that the heating was negligible at 10
and 100 AU.  The grain should therefore be able to reach the disk
largely unchanged \citep[e.g.][]{1994Icar..112..430R}.

We can therefore justify our simple treatment of dust injection.
Dust particles are added to our uppermost grid cell in the local disk
slice with an initial velocity of 0.  From that location, particles
will settle down and become part of the coagulation process.

\section{Results I: Effect of residual infall}\label{sec-results-1}
We completed the coagulation simulations at three different radii in the
disk: 1 AU, 10 AU and 100 AU. We assumed three different infall rates
(0, 10$^{-11}$, 10$^{-7}$)M$_{\odot}$/yr, which translate into local
infall rates in g cm$^{-2}$ s$^{-1}$ using
Eq.~(\ref{eq-local-trickling-rate}). We also assumed two different
strengths of turbulent mixing: a weakly turbulent disk
($\alpha=10^{-6}$, as a nearly laminar case) and a strongly turbulent
disk ($\alpha=10^{-2}$, as typically found by MRI instability
calculations, and also needed to explain the high accretion rates during the
early stages of disk evolution). Here, $\alpha$ is the Shakura-Sunyaev
parameter for viscosity, which can be used to parametrize the
strength of the turbulence, assuming a Stokes number of unity. We do
not include fully laminar conditions, since we believe that these are
unrealistic.  We allow each model to run for 1 Myr, in a few time-consuming
exceptions we stopped earlier if a steady state had already been
reached.  This provides a 3$\times$4$\times$3 matrix of models.

\begin{figure*}
\centerline{
\includegraphics[width=18cm]{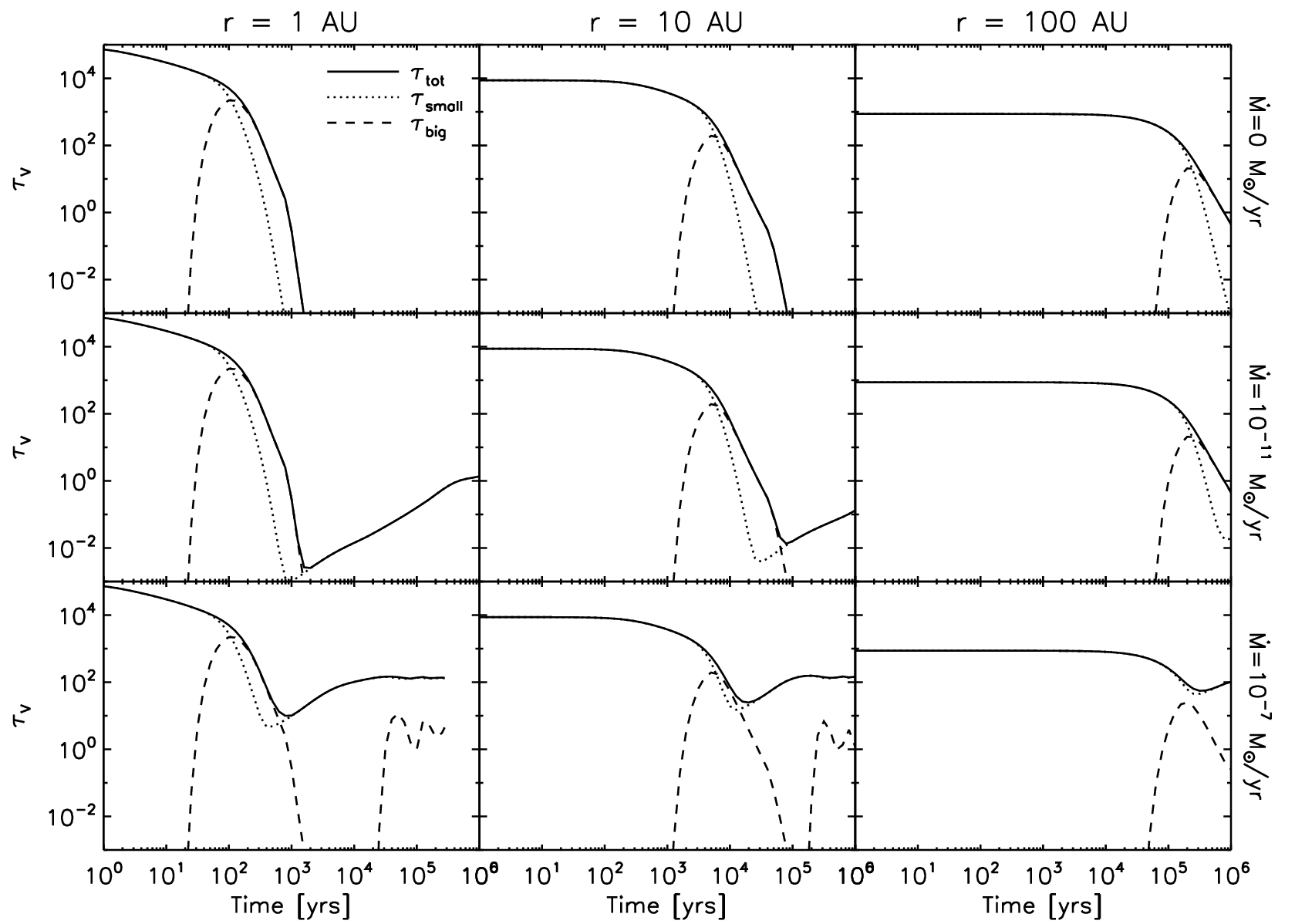}
}
\caption{\label{fig-tau-time-1d-2}Like fig.~\ref{fig-tau-time-1d-2}, but with
  with turbulent parameter $\alpha=10^{-6}$.}
\end{figure*}

In Fig.~\ref{fig-tau-time-1d-2}, the resulting time-dependent optical vertical
depths at $\lambda=0.55$ \um are shown for the case of $\alpha=10^{-6}$. 
\begin{figure*}
\centerline{
\includegraphics[width=18cm]{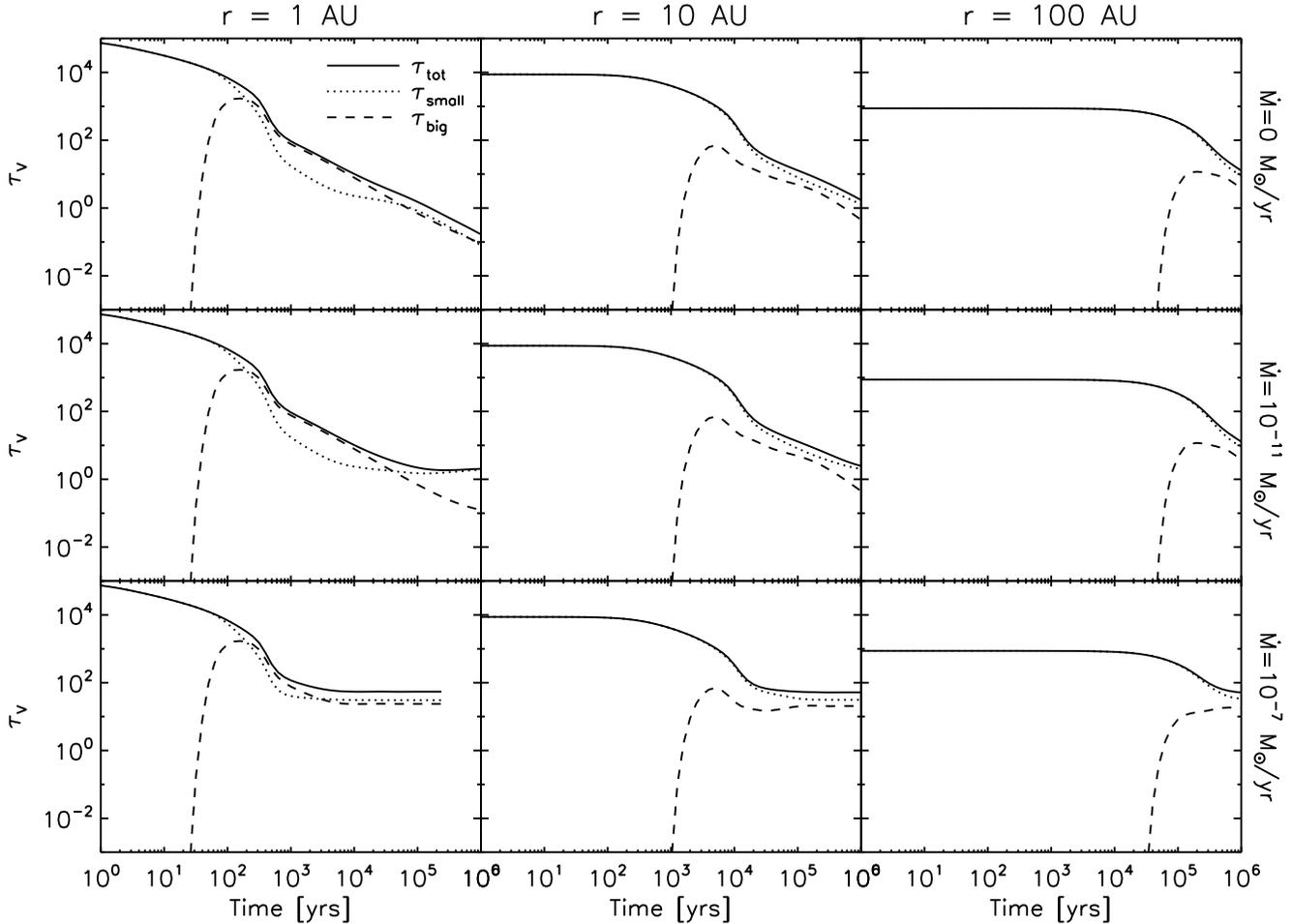}
}
\caption{\label{fig-tau-time-1d-6}The vertical V-band optical depth of the disk
with turbulent parameter $\alpha=10^{-2}$ at 1 AU, 10 AU and 100 AU (from
left to right) and infall rates 0, $10^{-11}M_{\odot}$/yr,
$10^{-7}M_{\odot}$/yr (from top to bottom). The dotted line is the
contribution to the optical depth by grains with radius $\le 3\mu$m, the
dashed line is the contribution by grains $>3 \mu$m.}
\end{figure*}
The top row shows the results without residual infall, which can be
most closely compared with the laminar models of DD05 and TIH05. It
should be noted, however, that the small, but non-zero, turbulence in
the present models has non-negligible effect on the results, and the
comparison is therefore only partly valid. In these top three models,
the coagulation is first driven by Brownian motion, which is
responsible for the weak reduction in optical depth between $t=$0 and
$t=$300 yrs for the R=1AU case. The optical depth then suddenly begins
to drop, as sedimentation-driven coagulation reaches its peak
intensity and efficiently removes small grains from the disk. The slow
further decline in optical depth is then caused by turbulence-driven
coagulation, which slowly removes approximately 0.1 percent fine dust
that escaped the sweep-up by settling grains. The final optical depth
at 1 Myr turns is 0.2 at 1 AU, 2 at 10 AU, and 12 at 100 AU. While the
outer regions of the disk are still optically thick, the inner regions
have become optically thin.  This was found by TIH05 to occur in their
laminar models and they proposed that this might be a possible
mechanism for explaining `inner holes' in certain disks
\citep[e.g.][]{2004ApJS..154..443F, 2003A&A...401..577B,
  2002ApJ...568.1008C}. The second row of Fig.~\ref{fig-tau-time-1d-6}
shows the same models, but with a continuous trickling infall at a
(global) rate of $\dot M=10^{-11}M_{\odot}/$yr. We see that at 1 AU
this causes the optical depth to become constant at a value of 2. The
outer disk is hardly affected. At an infall rate of
$\dot{M}=10^{-7}M_{\odot}/$yr, lower row, the optical depth at 1 AU
becomes constant at 80 and at 10 AU and 100 AU at 70. These values are
sufficiently high for the disk to be considered optically
thick disk.

Figure \ref{fig-tau-time-1d-2} shows the same models as
Fig.~\ref{fig-tau-time-1d-6} but now with $\alpha=10^{-2}$. Again, the
top row shows the models without infall.  As for the low-$\alpha$
case, Brownian motion produces an initially slow decline in the
optical depth up to the onset of sedimentation-driven
coagulation. However, in contrast to the low-$\alpha$ case, the
subsequent decline in optical depth is extremely fast, which differs
from our results in Fig.~\ref{fig-tau-time-1d-6}.  Turbulence-driven
coagulation is far stronger than in the $\alpha=10^{-6}$ case, and
becomes highly effective in removing small grains.  In these
simulations we emphasize that the optical depth at late times is due
primarily to particles larger than 3 $\mu$m, an effect that might be
artificial and in part caused by the neglect of grain drift, unlike
the low-$\alpha$ models.  In the second row, we see that at 1 AU and
10 AU the trickling infall of fresh matter leads to a slow recovery in
the optical depth, after its strong previous decline.  The optical
depth is now primarily controlled by small particles. At 1 AU, the
optical depth levels off at around 1, and at 10 AU becomes 0.1, which
is still optically thin, and lower than at 1 AU. At this radius, the
optical depth is still increasing after 1 Myr. At 100 AU the infall
has not produced any noticeable effect.  In the last row, we can see
that following the recuperation of the optical depth after its
reduction it appears to saturate at a value of 150 at 1 AU as well as
at 10 AU. The system reaches an equilibrium state, in which the rate
of fine dust removal appears to equal the rate of replenishment by
fresh fine dust. The equilibrium state is still imperfect, since the
large grain population continues to oscillate, as can be seen by the
dashed lines. However, since these large grains ($>3\mu$m) represent
only a small contribution to the optical depth, this is of no
relevance to the optical depth issue.  Even at 100 AU, we now see that
the replenishment by new grains is effective and increases the optical
depth, but an equilibrium state is not yet reached.

Judging from these results, we see that, in principle, even a 
small residual infall rate onto the disk can explain
the observed optical thickness of T Tauri star disks, even if
fragmentation is not active and therefore cannot replenish the
population of small grains.  Whether infall scenarios with these rates in
the late evolutionary stages are realistic is, however, still a major
issue of concern, which we discuss in Sect.~\ref{sec-discuss}.

\section{Results II: Effect of reduced initial fine-dust abundance}
\label{sec-results-2}
In the models of coagulation described here, it is seen that the dust
in the disk coagulates initially extremely rapidly, reducing the
optical depth to almost zero apart from that due to a trickling
accretion of new material.  It is striking, however, that even a very
low residual infall rate of, say, $10^{-11}M_\odot/$yr can have this
effect.  At first sight this appears difficult to understand, because
at this trickling rate it takes 5 Gyr to replenish fully a disk of
$0.05M_{\odot}$, whereas the removal of small grains occurs within
$10^4$ years. One would expect the replenishment to be unable to
compete with the coagulation.  The speed of coagulation is however
inversely proportional to the density of dust. Therefore, the time to
grow to sizes that no longer contribute to the optical depth will be
inversely proportional to the density of fine-grain
material. \textit{If we begin with more fine-grained dust, we
  therefore end with less.}

This suggests another way to retain the value the optical depth in a
disk for a long time.   Our simulations have consistently
started from an already constructed disk, i.e.~after the early build-up
phase of the disk. Without justification, we have{\em assumed} that
at this point in time the dust population is interstellar in both size
distribution and abundance. However, coagulation is also active during
the build-up phase of the disk
\citep{1999ApJ...524..857S,2001ApJ...551..461S}, and it is therefore
to be expected that by the time the main infall phase is over (and our
simulation begins) the abundance of fine-grain dust has already
decreased. 

\begin{figure}[t]
\includegraphics[width=9cm]{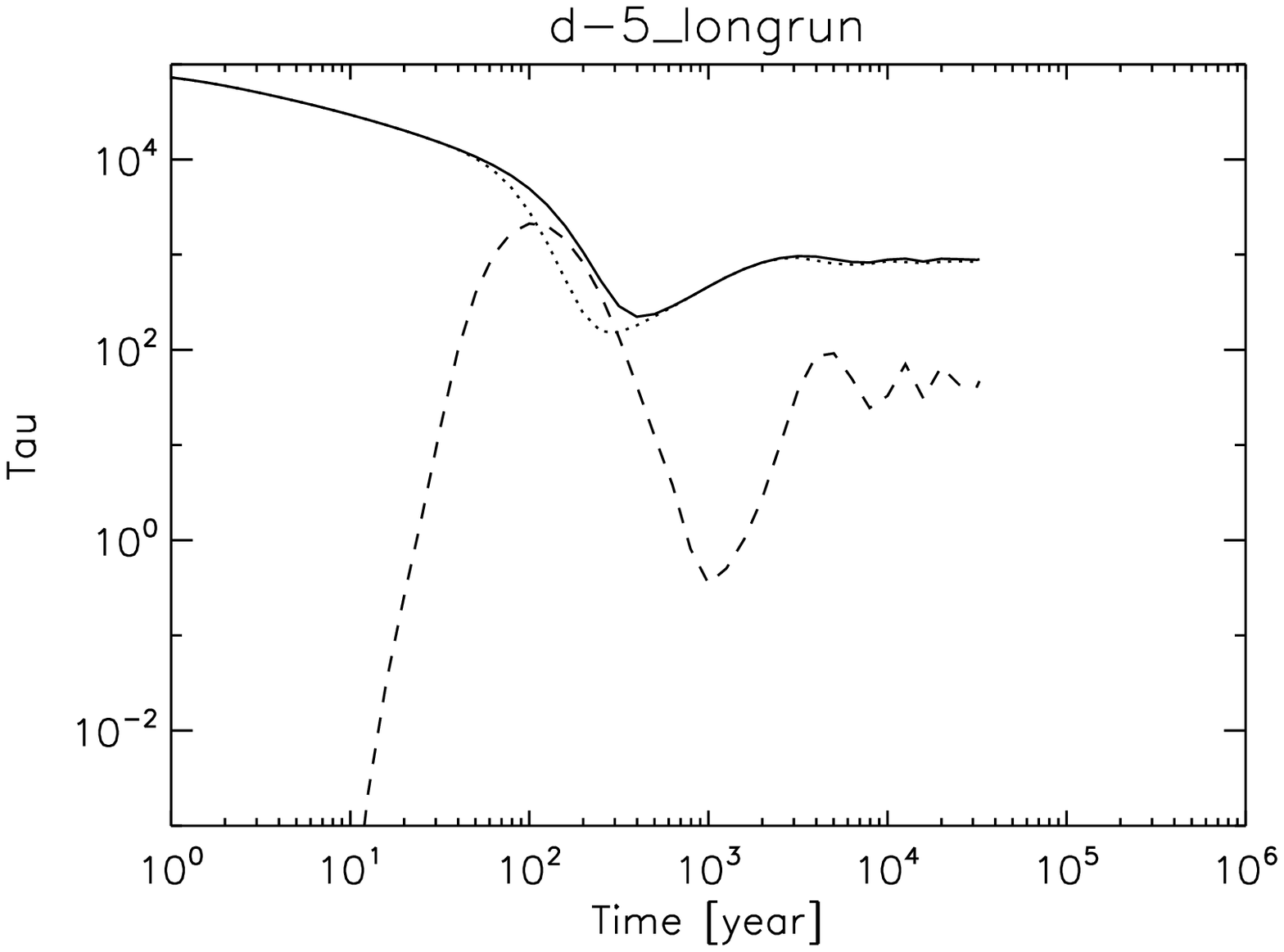}
\caption{\label{fig-high-infall}The time-dependent optical depth for the
case when we apply our model to the {\em build-up} phase of the disk, with
an infall rate of $10^{-5}M_{\odot}/$yr. Since this simulation is merely
meant as an `initial condition' for the Myr-long coagulation process studied
in this paper, we run it only until an equilibrium state has set in. As can
be seen in this figure, this happens after about $10^4$ years.}
\end{figure}

\begin{figure}[t]
\includegraphics[width=9cm]{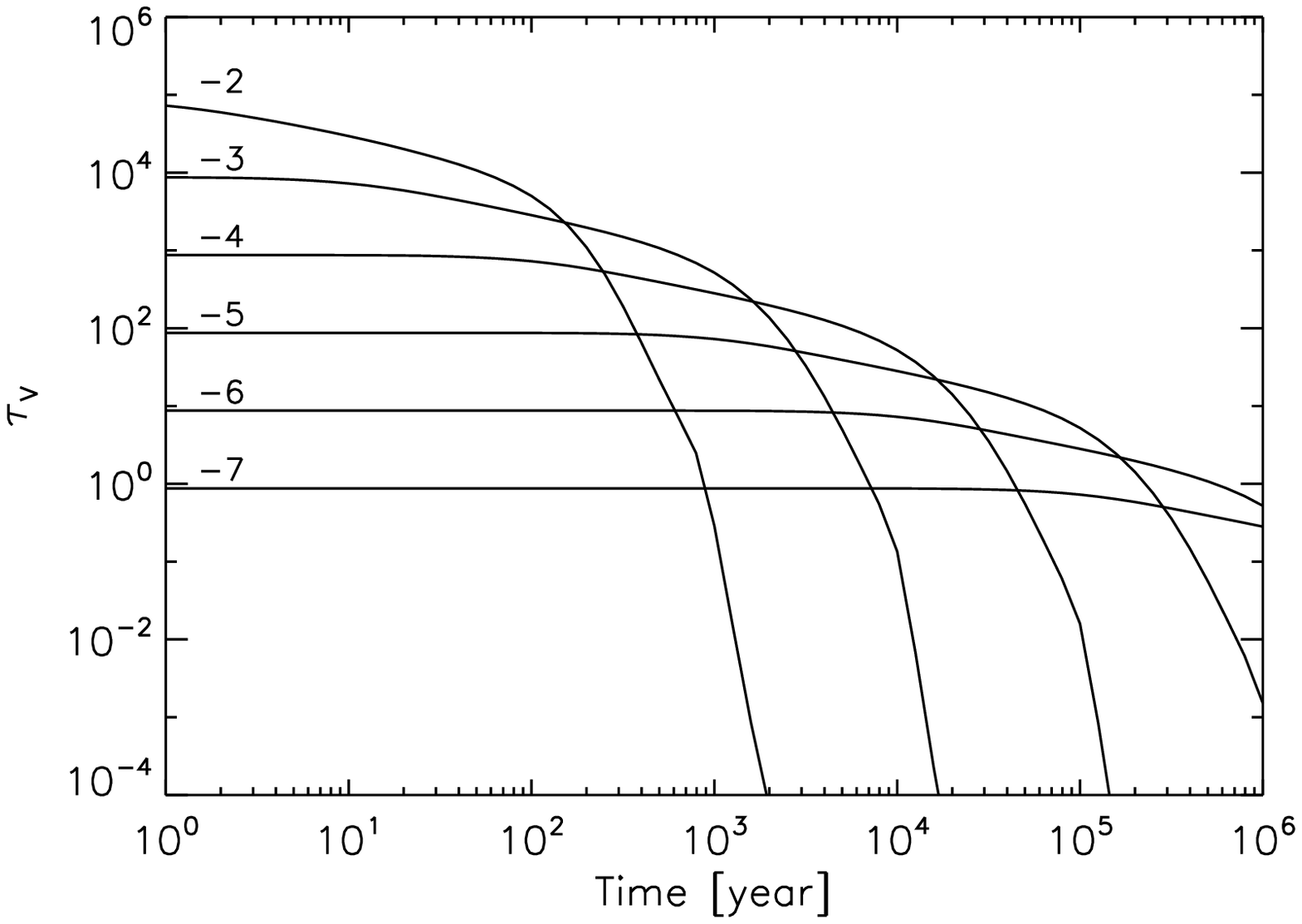}
\caption{\label{fig-low-dtg}The time-dependent optical depth for the for
  disks with varying initial dust-to-gas ratio.  The curves are labeled with
  $\log f_{\rm dg}$.  This figure clearly demonstrates the ``The higher
  you start, the lower you end'' effect caused by the non-linear behavior of
  the coagulation equation.}
\end{figure}

To investigate the significance of this reduction, we could apply the
models of Sect.~\ref{sec-results-1} to the build-up phase, and observe
the equilibrium state that is reached.  We simply need to adopt a far
higher infall rate, i.e.~a rate at which a star can be built in about
0.1 Myr.  In our case this corresponds to an infall rate of about
$10^{-5}M_{\odot}/$yr.  In Fig.~\ref{fig-high-infall}, we see that, at
these high infall rates, the equilibrium is reached at a very early
stage, justifying the use of this state as an initial condition.  In
this particular case the figure shows that the optical depth is
reduced by about 2 orders of magnitude compared to the
``interstellar'' dust abundance case.  As an initial condition of our
Myr-long simulation, it would therefore be justifiable to begin with a
fine-dust abundance that is a factor of 100 below the interstellar
medium value.  However, we have not modeled this buildup phase
consistently, and more sophisticated physical ingredients, such as
pre-collapse coagulation, disk accretion, and radial drift of larger
grains, would need to be included, which is well beyond the scope of
this paper.  We therefore study the effect of a reduced initial
fine-dust abundance on the Myr-long simulations, by using the initial
abundance as a free parameter.  This is implemented by completing
models without any residual infall, but with decreased initial
dust-to-gas ratio. We ignore any population of very large grains that
may have been produced during the build-up phase.

The results displayed in Fig.~\ref{fig-low-dtg} indicate that the
initial optical depth is proportional to the initial dust-to-gas ratio
(which is obvious) and that this optical depth can be sustained for a
time that is inversely proportional to the initial dust-to-gas ratio.
For example, with an interstellar dust-to-gas ratio (i.e.\ $10^{-2}$)
the optical depth is almost $10^5$ at first, but a steep decrease
occurs after 100 years.  If we begin the simulation, however, with a
dust-to-gas ratio of $10^{-7}$ (i.e.\ a factor of $10^5$ lower than
interstellar), then the initial optical depth is close to 1 and can be
sustained for about 1 Myr.  It therefore appears that the strong
reduction of optical depth may occur as late as 1 Myr, but only if
start with an optical depth that is already close to 1.

This implies that there exists a small region of parameter space in
which the disk can remain marginally optically thick ($\tau_V\simeq
1$) for 1 Myr, without fragmentation or any other means of
replenishment of small grains. However, many disks live even longer,
and there is a large volume of observational evidence suggesting that
these disks have higher optical depths (e.g. Forrest et al. (2004)
\nocite{2004ApJS..154..443F}, and Dullemond and Dominik (2004)
\nocite{2004A&A...417..159D}; for a review of fitting disk models to
observations, see Watson et al, (2007)\nocite{2007prpl.conf..523W}).

It will also be difficult to produce this specific initial state, for
the following reason: low optical depth equilibrium states require a
long time to be established. The equilibrium state reached during the
disk build-up phase is {\em defined} to be the state in which the
timescale of replenishment of fine dust equals the timescale of
removal of fine dust by coagulation. Therefore, the build-up time of
the state {\em with} replenishment equals the lifetime of the state
{\em without} replenishment.  For example, during the disk build-up
phase (clearly at a low infall rate) we would require 1 Myr to achieve
an equilibrium state that could, after the end of infall, be sustained
for another 1 Myr as an optically thick disk.  It is believed,
however, that the build-up phase of a star+disk system (i.e.\ the
class O and class I phases in the Lada/Andre scheme) has a duration of
less than 1 Myr \citep{1994ApJ...420..837A, 1994ASPC...65..197B},
implying that this phase is significantly shorter than the typical
disk lifetime.  The conclusion is that the dust {\em cannot} reach a
steady state that lasts for a few Myrs. Any steady state reached
during this time is certain to decay shortly after the main infall
phase.  It is also possible that no steady state of the dust
population is reached at all during this disk build-up phase.  Both
situations are not promising for keeping the optical depth high over a
few Myrs time.

\section{Discussion}
\label{sec-discuss}

\subsection{Consequences for dust fragmentation}
\label{sec:cons-dust-fragm}

In Sect.~\ref{sec-results-2}, we have shown that a low value of
fine-grained dust density in the disk can in principle persist for a
long time, but that there is only a narrow range in parameter space
for which the disk optical depth remains above unity for a few million
years.  It also takes a long time to reach this long-lasting state.
This result is fully independent of the mechanism that produces the
small grains in the disk.  We can also apply it to small grains
produced by aggregate fragmentation and erosion (instead of trickling
infall). If, after studying possible alternatives in this paper, we
return to fragmentation and erosion as the most likely sources of
small grain replenishment in disks, then we must conclude that these
must be ongoing processes lasting throughout the few-million-year
lifetime of the disk.  If this replenishment process were to switch
off for some reason, the disk would quickly become
optically thin.

\subsection{Limits on infall onto T Tauri star disks}
\label{sec:limits-infall-onto}

How realistic is the assumption of low-level infall onto the disk?
There are several ways to place constraints using both observational
and theoretical arguments.  In the following, we show that
acceptable infall rates are constrained to be in the  narrow range
between 10$^{-7}$ and 10$^{-8}$M$_{\odot}$/yr.

\subsubsection{Optical depth of the envelope}
\label{sec:optic-depth-envel}
We emphasize hat we consider sources with only limited extinction
toward the star.  Many of the sources with optically thick disks have
circumstellar reddening $A_V$ lower than (sometimes far lower than)
two \citep[e.g.][]{1997A&A...324L..33V}.  The largest contribution to
the optical depth of the infalling envelope originates in the region
around $r\simeq r_{\rm centr}$.  Using the density distribution from
the Ulrich infall model, we arrive at an optical depth in the visual
of $1.88$ for an infall rate of $4\times 10^{-7}$M$_{\odot}$/yr.
Therefore, the maximum infall rate that is consistent with the class
II status of these sources is about this value.  The optical depth at
different infall rates obviously scales linearly with the infall rate.

\subsubsection{Stellar winds}
\label{sec:stellar-winds}
Young stars are also known to possess stellar winds, which can remove
residual circumstellar envelopes.
Using infrared and UV lines, \citet{2005ApJ...625L.131D} 
  argued that TW Hya accelerates a hot wind with a velocity of 400km/s
  and a mass loss rate of about 10$^{-11}$M$_{\odot}$/yr, which should
  be able to quench any infall with rates below
  10$^{-9}$M$_{\odot}$/yr.   The evidence for this
  wind was however questioned by \citet{2007ApJ...655..345J}.  If
  these  wind properties were typical for T Tauri stars,
  10$^{-9}$M$_{\odot}$/yr would be a lower limit for residual infall
  to reach the disk.


\subsubsection{Radiation pressure and dust/gas separation}
\label{sec:radi-press-dustg}
Another obstacle to any residual infall is radiation pressure.  At
low infall rates, the envelope is optically thin and the radiation 
reaches all the dust grains in the infalling gas.  The problem of
radiative blowout of grains from main-sequence stars was
studied extensively in the context of the Vega phenomenon and for
$\beta$ meteorites in the Solar System \citep{1979Icar...40....1B,
  1980IAUS...90..293G, 1988ApJ...335L..79A, 2000ApJ...539..424K}.  For
the luminosity of the Sun, a small range of dust particle sizes around
0.1\um have the correct properties for radiation pressure to exceed
gravitational forces \citep{1979Icar...40....1B}.  For more luminous
sources such as A stars or even their still more luminous Herbig Ae
precursors, the blowout limit is pushed to 10\um and beyond, such  that
grains in a size range from 0.01 to 10\um should be blown away from the
star \citep{1988ApJ...335L..79A}.  The one difference in the scenario
considered here is of course that the dust grains are embedded in an
infalling gas envelope, which is not affected by any significant radiation
pressure and could possibly drag the dust particles along.

The drift velocity caused by the radiation pressure
on dust grains must then exceed the infall velocity of the material.  We
can place a limit on the mass accretion rates necessary for dragging the
grains along in the following way.  In a spherical free-fall accretion
scenario with a constant infall rate $\dot{M}$, the local velocity is
given by $v(r)=-\sqrt{2GM_{\star}/r}$ and the gas density is given by
$\rho=\dot{M}/(4\pi r^2 |v|)$.  We assume that  a dust
grain is embedded in that infall.  The stellar radiation in incident
on the grain with a flux $L_{\star}/(4\pi r^2)$.  The radiation force on
the grain is then
\begin{equation}
\label{eq:5}
F_{\rm rad} = \frac{L_{\star}}{4\pi r^2c}Q_{\rm pr}\pi a^{2}
\end{equation}
where $a$ is the grain radius and $c$ is the speed of light. The
efficiency for pressure transmission to the grain is $Q_{\rm pr}=Q_{\rm
  abs}+Q_{\rm scat}\langle\cos \alpha\rangle$ is , with $Q_{\rm abs}$
being the absorption efficiency, $Q_{\rm scat}$ the scattering
efficiency, and $\langle\cos \alpha\rangle$ the angular average of the
scattering phase function \citep{bh83}.  $Q_{\rm pr}$ is dependent on
the grain size.  In Herbig AeBe stars with a radiation temperature of
10000\,K, $Q_{\rm pr}$ will be of order 1 for all grain sizes
considered here, whereas 
in the cooler radiation field of T Tauri stars, it will be of order 1
for $a\ge 1 \um$, and somewhat smaller for 0.1\um grains.  The grains
will not move toward the star if the radiation force can at least
balance the drag force of the gas streaming toward the star.  The
limiting case is reached if the grain is at rest with respect to the
star.  Then the gas streams by at a velocity $v=v_{\rm ff}$.
As the free-fall flow is supersonic, we must apply the friction law
for the supersonic case in which we have
\begin{equation}
\label{eq:4}
F_{\rm fric}=-\pi a^2\rho v^2 \quad.
\end{equation}
To keep the grain fixed in space, the friction force must be balanced
by the radiation force,
\begin{equation}
\label{eq:6}
F_{\rm fric}+F_{\rm rad}=0
\end{equation}
\noindent
where we are neglecting the gravitational force on the dust grains,
restricting the discussion to grains that are clearly inside the
blow-out size regime.  This equality is satisfied for a critical mass
accretion rate
\begin{align}
\label{eq:8}
\dot{M}_{\rm cr} &= \frac{3L_{\star}Q_{\rm
    pr}\sqrt{r}}{\sqrt{2GM_{\star}}c}\\
&= 4.8\times 10^{-8}\frac{M_{\odot}}{\mbox{yr}}
\frac{L_{\star}}{L_{\odot}}
\frac{Q_{\rm pr}}{1}
\sqrt{\frac{M_{\odot}}{M_{\star}}}
\sqrt{\frac{r}{10000\mbox{AU}}}
\end{align}
As we can see, the mass loss rate required to drag dust grains along
is a function of distance from the star.  Far away from the star, a
relatively high mass infall rate will be required for the dust grains
to be dragged along.  As the infall moves closer to the star, it
becomes easier to drag the dust grains along.  Therefore, the true
limiting infall rate is determined at large distances, where the free
fall toward the star begins.  If the infall starts at 10000\,AU from
the star, we find a minimum infall rate of
$4.8\times10^{-8}$M$_{\odot}$/yr for solar parameters.

It is therefore clear that gas drag may help to overcome the outward
pushing radiation force only for the highest infall rates.  For Herbig
AeBe stars, the infall rates resulting from our estimate contradict
directly the observations of low $A_V$, and imply that low-level
infall cannot play a role.  For T Tauri stars, there is a window of
infall rates between approximately $4\times10^{-7}$ and
$5\times10^{-8}$M$_{\odot}$/yr, which in principle would allow
accretion of dust grains onto the disk.

\subsection{General remarks}

In this paper we have studied two scenarios to keep the disk optically thick
in spite of dust coagulation.

The first scenario is the low rate of matter infall onto the
disk.  This could help to retain a sufficiently high optical depth.
However, it is unclear if these infall rates are in fact present.  We
can place severe limits on the allowed rates by considering  the optical
depth of the infalling envelope, and the effects of stellar winds
and radiation pressure on dust (see
Sect.~\ref{sec:limits-infall-onto}).

The second scenario involved beginning the coagulation simulation with
a reduced abundance of fine dust. We have justified these special
initial conditions with ongoing coagulation during disk
build-up. However, we have concluded that the special initial state
needed to sustain $\tau_V\gtrsim 1$ without dust replenishment cannot
be achieved in the time available.  Even if this were achievable, the
optical depth that one can sustain for 1 Myr does not exceed
$\tau_V\simeq 1$, in contradiction with observational evidence.

Therefore, fragmentation and erosion of aggregates appear still to be
the most promising explanation for the observed small dust grain
populations in disks of 1 Myr age or older, as proposed by DD05.

It should be noted that the models we have presented in this paper are
based on considerable simplifications. The most important of these is
that we do not include the accretion of gas in the disk, toward the
star.  Any material that is located at 1 AU will be advected toward
the star on a viscous timescale, which for a turbulence of
$\alpha=10^{-2}$ (assuming that the Schmidt number is 1) equals $10^4$
years, and the structure of the disk may evolve significantly over a
period of 1 Myr.  As stated before, we have not included radial drift
of grains.  It is clear that the next step will have to be
the full time-dependent modeling of the coupled equations of disk
formation and evolution, and dust coagulation. With such models one
could repeat the study completed here in a fully consistent way.

\section{Conclusion}

With this study, we arrive at the following conclusions:

\begin{enumerate}

\item\label{item:1} Low-level infall onto a disk can have a
  surprisingly strong influence on the steady-state abundance of small
  dust grains in a protoplanetary disk, in particular if the infall
  reaches a disk that is initially cleared of small grains.  This is
  due to a strong non-linear effect:  the high initial dust content of
  the disk leads to very fast coagulation and a complete loss of small
  grains.  The low dust densities produced by constant trickling are
  balanced by a far slower coagulation.


\item\label{item:3}  To be significant, the trickling rate must
  exceed 10$^{-11}$M$_{\odot}$/yr.  Over a lifetime of 1\,Myr,
  this corresponds to a total trickling mass in gas of
  10$^{-5}$M$_{\odot}$.

\item\label{item:4} In our calculations, the optical depth of the disk
  is produced by grains smaller than 3\um.  In this case, we expect
  that the 10\um feature, an important indicator for small grains,
  will be present.  Radiative transfer calculations will be required
  to determine whether this is the case.

\item\label{item:5} Almost perfectly laminar disks can remain
  marginally optically thick even without trickling, but for these
  disks the optical depth is produced by \emph{large grains}
  ($>3\um$).

\item\label{item:6} Radiation pressure in Herbig stars excludes
  trickling as a viable mechanism for keeping the disk optically
  thick.  Even for T Tauri stars, mass infall rates higher than
  10$^{-8}$M$_{\odot}$/yr are needed to overcome the effects of
  radiation pressure on small grains.  Large grains (above the
  blow-out limit) can be accreted more easily.  For Herbig stars, the
  size limit of the order of $a\ga 10\um$, rendering it useless.  In T
  Tauri stars, we find $a\ga 1\um$, which is dependent on luminosity.

\item\label{item:7} Low initial dust-to-gas ratios could lead to disks
  with long-lasting optical depth of order unity.  However, it appears
  unlikely that these states correspond to initial conditions because
  the time required to achieve these states is comparable to their
  duration.

\end{enumerate}

The above conclusions can be summarized in the following way: {\em
  While a low-level trickling infall of residual envelope matter onto
  the disk might conceivably sustain the disks's optical depth above
  unity even without grain fragmentation, the conditions for this
  residual infall are very stringent.  Ongoing and perpetual
  aggregate fragmentation and erosion in the disk as an explanation
  for the disk's optical depth remains a more effective explanation.}

\begin{acknowledgements} CPD acknowledges financial support from the
  Max Planck Gesellschaft under the SNWG program. CD would like to
  thank the Dutch Top Research School NOVA (network 2) for financial
  support and the Leids Kerkhoven-Bosscha Fonds for travel support.
  We thank G. Wurm for a discussion about the effect of initial dust
  densities which has in part triggered this work.
\end{acknowledgements}

\bibliographystyle{apj.bst}

\end{document}